\documentclass[twocolumn]{aastex63}
\begin{document}
\shorttitle{An Intruder Star in the Hyades}

\shortauthors{Hu, Q.S. et al.}

\title{A Low-speed Intruder Star in Hyades: A Temporary Residence}

\author{Qingshun Hu}
\affil{Xinjiang Astronomical Observatory, Chinese Academy of Sciences, Urumqi, Xinjiang 830011, People's Republic of China}
\affil{University of Chinese Academy of Sciences, Beijing 100049, People's Republic of China}

\author{Yu Zhang}
\email{zhy@xao.ac.cn}
\affil{Xinjiang Astronomical Observatory, Chinese Academy of Sciences, Urumqi, Xinjiang 830011, People's Republic of China}
\affil{University of Chinese Academy of Sciences, Beijing 100049, People's Republic of China}

\author{Ali Esamdin}
\affil{Xinjiang Astronomical Observatory, Chinese Academy of Sciences, Urumqi, Xinjiang 830011, People's Republic of China}
\affil{University of Chinese Academy of Sciences, Beijing 100049, People's Republic of China}

\author{Dengkai Jiang}
\affil{Yunnan Observatories, Chinese Academy of Sciences, Kunming 650216, People's Republic of China}

\author{Mingfeng Qin}
\affil{Xinjiang Astronomical Observatory, Chinese Academy of Sciences, Urumqi, Xinjiang 830011, People's Republic of China}
\affil{University of Chinese Academy of Sciences, Beijing 100049, People's Republic of China}

\author{Ning Chang}
\affil{Xinjiang Astronomical Observatory, Chinese Academy of Sciences, Urumqi, Xinjiang 830011, People's Republic of China}
\affil{University of Chinese Academy of Sciences, Beijing 100049, People's Republic of China}

\author{Haozhi Wang}
\affil{Xinjiang Astronomical Observatory, Chinese Academy of Sciences, Urumqi, Xinjiang 830011, People's Republic of China}
\affil{University of Chinese Academy of Sciences, Beijing 100049, People's Republic of China}

\begin{abstract}

We hereby report a low-speed (about~21~km$\cdot$~s$^{-1}$ with respect to the Sun) intruder member in the Hyades cluster based on the data in the literature. The results show that the star is a nonnative member star for the Hyades, with its radial velocity being smaller than the radial velocity of the Hyades cluster, even exceeding the standard deviation of the radial velocity of the cluster by a factor of 9. Furthermore, by analyzing and comparing the orbits of this star and its host, it may have intruded into its host in the past 2~Myr. If the star's current motion orbit remains unchanged, it may leave its host in the next 2~Myr. This implies that the intruder star may be temporarily residing in the cluster. This study presents the first observational evidence of a star intrusion into a cluster, which suggests that more evidence may be found.

\end{abstract}


\section{Introduction}

Members of an open cluster are usually born in the same molecular cloud at almost the same time. So they have roughly the same age, metal abundance, and similar kinematic characteristics. And they generally move together in the Milky Way as a comoving group. During the evolution of the cluster, its member stars slowly break away from it and enter its tidal tail structure \citep[e.g.][]{tang19, zhang20}, even merging directly into the field stars.

Some early-type stars escaping from clusters or associations to come into being field stars have been studied by several works \citep[e.g.][]{gies87, dewi04, dewi05}. However, the escape of late-type stars from clusters or the capture of late-type stars by clusters has not been observationally paid much attention or is even a gap, especially the capture of late-type stars by clusters. In most cases, the parameters of late-type stars are unknown or uncertain, because they are too faint to be observed accurately. Furthermore, nonnative late-type member stars enter the interior of a cluster during its motion. Such a phenomenon is theoretically possible, but there is no evidence presented for this phenomenon in actual observations. In addition, most clusters are usually relatively distant, which poses a challenge for finding such observational evidence. It is expected that such observational evidence can help us to modify or constrain the evolutionary models of clusters and also improve our understanding of many interesting open questions, such as the dynamical evolution of cluster members \citep[e.g.][]{port10, kuhn14}.

Based on the data from the literature, we stumbled upon a late-type intruder star in the Hyades cluster when studying the morphological evolution \citep[e.g.][]{Hu21a, Hu21b} of clusters. Accordingly, we have investigated the relationship between this star and the Hyades in this work. The rest of this article is structured as follows: Sect.~2 describes the data filtering process; Sect.~3 presents the member's analysis and motion orbit analysis for the target star. Two interesting points are discussed in Sect.~4, and conclusions are presented in Sect.~5.

\section{Data}

\label{data}

The member stars of the Hyades (Melotte~25) in this work were taken from \citet{cant20} who published a members' catalog of 2017 open clusters with DR2 data, which is, the update of the 1481 clusters from \cite{cant20b}. They released all cluster members with probabilities ranging from 0.7 to 1.0, which means that the members are relatively reliable. In their catalog, we found that the Hyades contained 515 members. We obtained the DR3 data of those sources  (hereafter, \textit{Sample a}) via the available crossmatch table, and removed their DR2 data, among which there are only 279 stars with radial velocities. With the parallax cut below, only 271 stars with radial velocities (called \textit{Sample b}) remain. Meanwhile, all stars in the \textit{Sample a} and \textit{Sample b} have a member star probability of 1. Owing to the high probability of these stars, our work is focused on the \textit{Sample a} and \textit{Sample b}.

\begin{figure*}[ht]
	\centering
	\includegraphics[width=160mm]{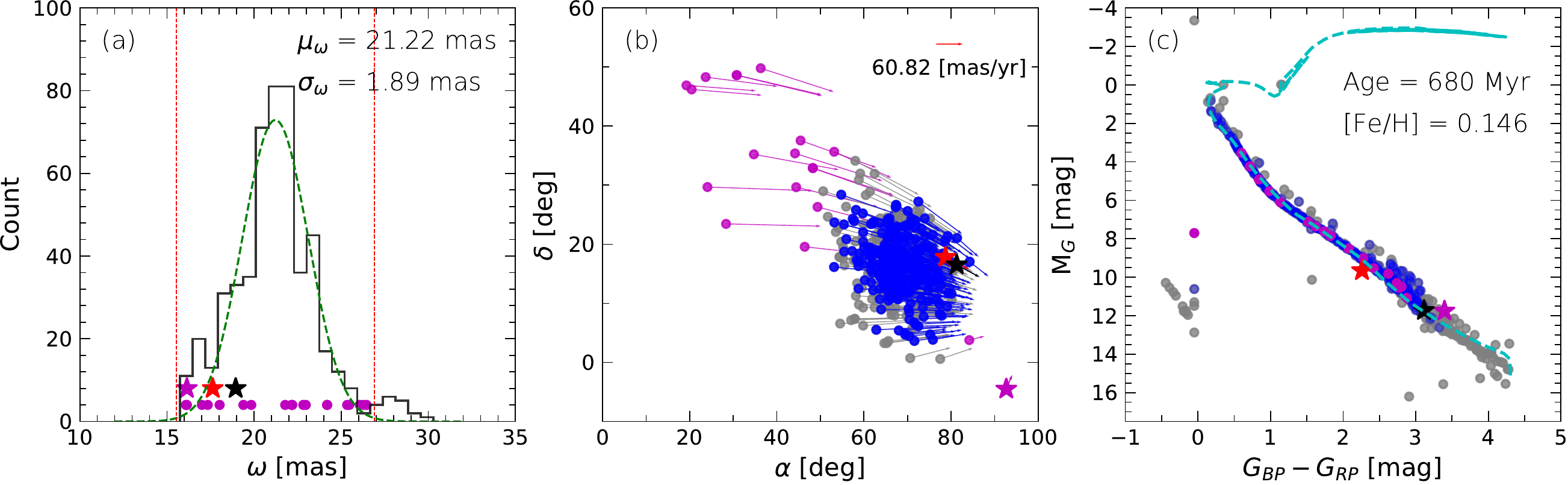}
	\caption{Comparison in multidimensional parameter space. Members in \textit{Sample a} published by \citet{cant20} are shown as gray dots. In blue dots, we plot the stars in \textit{Sample b}, with the \texttt{H-star} displayed as a red pentagram, along with the suspected intruders marked by the black pentagram. Stars in \textit{Sample c}, plotted in purple dots, are within the tidal structure of Hyades released by \citet{Jera21}, along with the suspected intruders marked by the purple pentagram. (a): Histogram of parallax ($\omega$) for \textit{Sample a} (black line). The \texttt{H-star}, the two suspected intruders, and \textit{Sample c} are located in the area of 3$\sigma_{\omega}$ boundary (red dashed lines), which is based on the use of a single Gaussian profile (green dashed line) for this histogram. (b): Spatial distribution for the Hyades members (\textit{Sample a}, \textit{Sample b}, and \textit{Sample c}), the \texttt{H-star}, and the two suspected intruders with their proper motion vectors. Arrows point out the direction of proper motion vectors for each star and arrow length is in proportion to the proper motion. (c): Color-magnitude diagram of Hyades cluster (\textit{Sample a}, \textit{Sample b}, \textit{Sample c}, the \texttt{H-star}, and two suspected intruders), with the isochrone plotted in cyan dashed line.}
	\label{Melotte25_Parameters_position}
\end{figure*}

We note that although the stars in \textit{Sample a} are more reliable, they only cover the main (core) region of this cluster instead of its tidal tail structures; see gray dots in Fig~\ref{Melotte25_Parameters_position}. The tidal tails of this cluster have been detected by several works \citep[e.g.][]{rose19, Jera21}. We hope to find some reliable, member stars distributed in the tidal tails of the Hyades. These stars can be used to corroborate or verify if our target star is an intruder. For Hyades, \citet{Jera21} detected its tidal tail structures and released a total of 1109 members by combining the Gaia DR2 data with the kinematics simulation of the Hyades cluster and using a new compact convergent point method. By crossmatching data, the DR3 data of these 1109 stars also were obtained. We also removed their DR2 data. The 1109 members not only appear in the core region of Hyades but are also located in its tidal tails. By removing the stars that duplicate \textit{Sample a} out of these 1109 stars, we obtained the possible member stars in the tidal tail of this cluster, which are 665 stars in total. Since their membership probabilities are not available, we set filtering conditions to filter these 665 stars and use the filtered data as supplemental data for \textit{Sample a}. The filtering conditions are shown below.

\begin{itemize}
\centering	
\item $\omega$ within $\mu_{\omega}$ $\pm$ 3$\times$$\sigma_{\omega}$;
\item $Rv$/$e_{Rv}$ $\geq$ 10;
\item $Rv$ out $\mu_{Rv}$ $\pm$ 3$\times$$\sigma_{Rv}$.
\end{itemize}

The $\mu_{\omega}$ and $\mu_{Rv}$ denote the Gaussian fitted means of the $\omega$ and $Rv$, respectively, with $\sigma_{\omega}$ and $\sigma_{Rv}$ representing their standard deviations. The $\mu_{\omega}$, $\mu_{Rv}$, $\sigma_{\omega}$ and $\sigma_{Rv}$ were provided in the (a) panels of Fig~\ref{Melotte25_Parameters_position} and Fig~\ref{Melotte25_Parameters_velocity}. The $e_{Rv}$ refers to the error of the radial velocity. With this filtering method, we got 19 stars, forming \textit{Sample c}, plotted in purple dots in Fig~\ref{Melotte25_Parameters_position} and Fig~\ref{Melotte25_Parameters_velocity}.

\section{Results}

\subsection{Members Analysis}

\label{Intruder}

\begin{figure*}[ht]
	\centering
	\includegraphics[width=160mm]{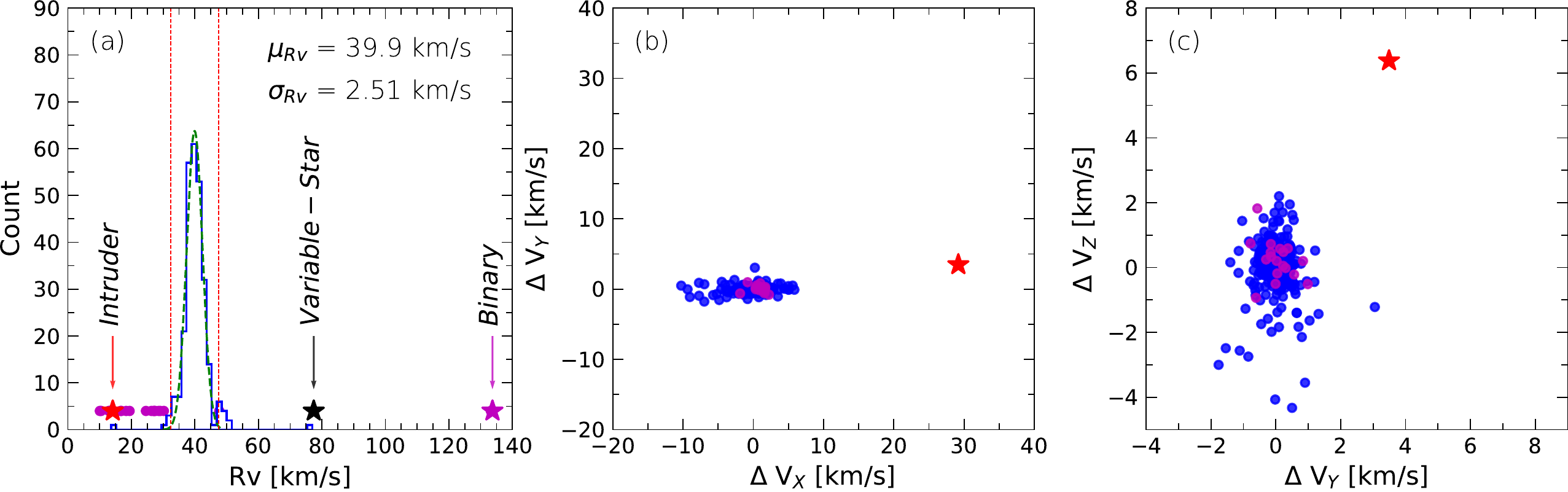}
	\caption{Comparison in velocity parameter space. All symbols are the same as that in Fig.~\ref{Melotte25_Parameters_position}. (a): Histogram of radial velocities for \textit{Sample b} plotted as the blue dots. The target star and \textit{Sample c} marked by the purple dots are out of the area of 3$\sigma_{Rv}$ boundary (red dashed lines), as well as two suspected intruders marked by the black and purple pentagrams, respectively. (b) and (c): scatter distributions for those with the radial velocities in $\Delta$V$_{X}$-$\Delta$V$_{Y}$ and $\Delta$V$_{Y}$-$\Delta$V$_{Z}$ velocity spaces, respectively. Note that $\Delta$V$_{X}$, $\Delta$V$_{Y}$ and $\Delta$V$_{Z}$ are relative Galactocentric velocities.}
	\label{Melotte25_Parameters_velocity}
\end{figure*}

We found an intruder member in the Hyades when analyzing the morphological evolution of this cluster. This intruder member was included in both \textit{Sample a} and \textit{Sample b}. We temporarily named it as `\texttt{H-star}', plotted in red pentagram, shown in Fig.~\ref{Melotte25_Parameters_position} and Fig.~\ref{Melotte25_Parameters_velocity}. In the two figures, we display the coincidences and distinctions between the \texttt{H-star} and its host cluster in multidimensional parameter and velocity spaces. It is obvious that its two-dimensional spatial position and proper motion are almost identical to those of the host, see the (a) subplot of Fig.~\ref{Melotte25_Parameters_position}. Besides, the \texttt{H-star} is also within the parallax range of the host cluster, which can be seen in the (b) panel of this figure. These are the reasons why it has been selected as a member of the host cluster many times \citep[e.g.][]{gaia18, rose19, cant20, freu20, Jera21}.

In \textit{Sample b}, we also found a suspected intruder star (plotted in black pentagram, shown in Fig.~\ref{Melotte25_Parameters_position} and Fig.~\ref{Melotte25_Parameters_velocity}), while in \textit{Sample c}, there is also a suspected intruder star (plotted in purple pentagram, shown in Fig.~\ref{Melotte25_Parameters_position} and Fig.~\ref{Melotte25_Parameters_velocity}). The locations of the \texttt{H-star} and the two suspected intruders in the color-magnitude diagram (CMD) of the host cluster are marked in the (c) subfigure of Fig.~\ref{Melotte25_Parameters_position}. We adopted the host cluster's parameters (Age = $\sim$680~Myr, [Fe/H] = 0.146, and Av = 0.093) provided by \citet{goss18} to obtain the isochrone (cyan solid line in the (c) subfigure) from PARSEC and overplotted it in the CMD. The \texttt{H-star} seems to appear on the lower edge of the main-sequence of Hyades. For the two suspected intruders above, the suspected intruder marked by the purple pentagram locates in the binary Main sequence of the Hyades, implying it should be a binary. Besides, although the one marked by the black pentagram is not in the main-sequence binary belt, we cannot determine whether it is a binary only by CMD. This is because binaries with a low mass ratio may also be in the position marked by the black pentagram. From here we do not see the age difference between the \texttt{H-star} and the cluster, along with the two suspected intruders, and it is excellent to seek the difference between their metal abundances.

For the two suspected intruders, we could not find information on their metal abundance. The spectrum of the \texttt{H-star} can be found in the fifth data release of LAMOST\footnote{http://www.lamost.org/lmusers/.} (LAMOST~DR5). But, the metal abundance ([Fe/H]) of the \texttt{H-star} is not available, as well as its radial velocity, due to the low signal-to-noise ratio (snrg = 18.03) of its observations. But, its spectral type was determined to be M-type in LAMOST~DR5. And, \citet{xian19} published the metallicity [Fe/H] = -1.271~$\pm$~0.044~dex for the \texttt{H-star} based on LAMOST DR5 with a modeling approach. 

\begin{figure*}[ht]
	\centering
	\includegraphics[width=100mm]{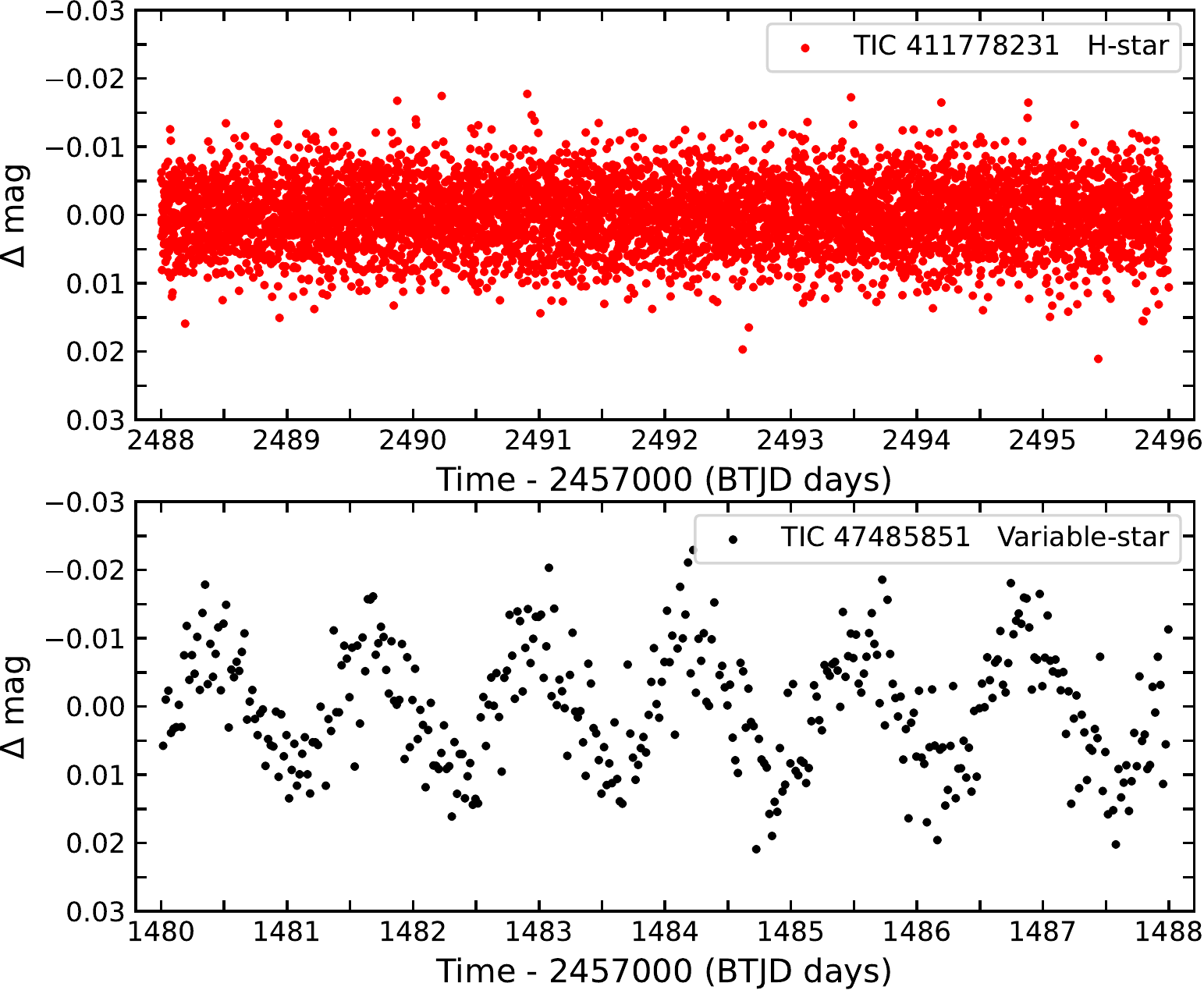}
	\caption{Upper panel: photometric variation for the \texttt{H-star}. Bottom panel: photometric variation for the suspected intruder marked by the black pentagram. The object names are labeled on the diagram.}
	\label{Lightcurve}
\end{figure*}

In general, the overall metal abundance of a cluster can be obtained in light of high-resolution spectroscopic data of its bright stars because of their high signal-to-noise ratio. \citet{made13} and \citet{cumm17} reported Hyades cluster averages of [Fe/H] = +0.130~$\pm$~0.009 based on 22 stars with the high-resolution spectral data and [Fe/H] = +0.146~$\pm$~0.004 based on 37 stars with the high-resolution spectral data, respectively. \citet{liu16} surveyed the metal abundance of the Hyades based on high-resolution, high signal-to-noise ratio spectra data, with [Fe/H] ranging from +0.115~$\pm$~0.016~dex to +0.203~$\pm$~0.015~dex. The metal abundance ([Fe/H]) of the \texttt{H-star} is significantly different from that of its host, implying that the \texttt{H-star} is not a native member of the cluster. The detailed parameters of the \texttt{H-star} are listed in Table~\ref{table:data}.

\begin{figure}[ht]
	\centering
	\includegraphics[width=56mm]{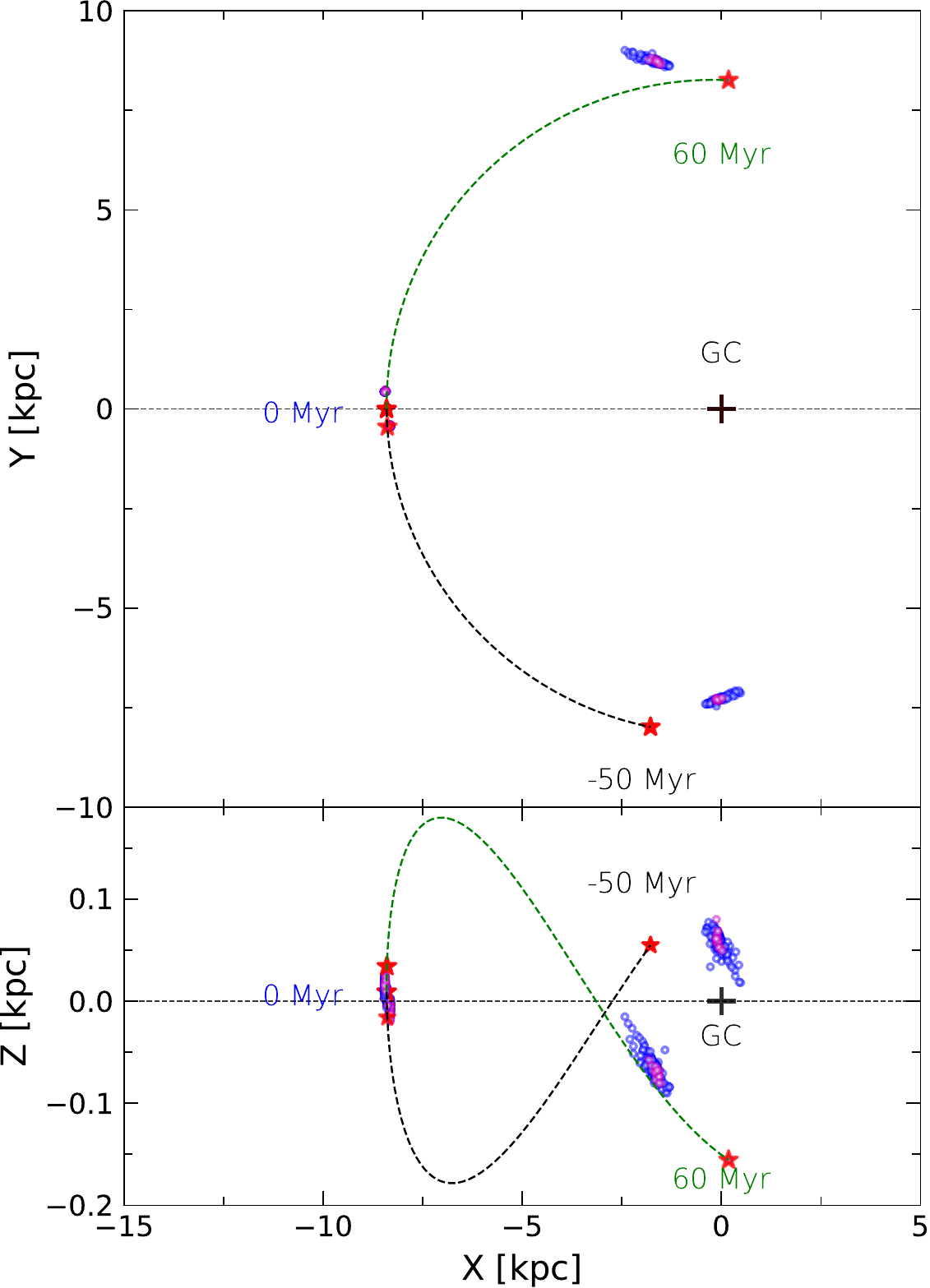}
	\caption{The trajectories of the \texttt{H-star} on different projection planes of the Galactic disk and the projection distributions of Hyades's members at some given moments (t = -50~Myr, -2~Myr, 0~Myr, 2~Myr, and 60~Myr). Upper panel: the projected trajectory of the \texttt{H-star} and the distributions of the Hyades's members in the X-Y plane. Bottom panel: the projected trajectory of the \texttt{H-star} and the distributions of the Hyades' members in the X-Z plane. Colored dots are the same as Fig.~\ref{Melotte25_Parameters_position}. The trajectories in black and green dashed lines represent the past 50~Myr and the forward 60~Myr, respectively.}
	\label{Melotte25_Orbits}
\end{figure}

Generally speaking, the member stars of a cluster should have approximately the same kinematic characteristics, for example, their radial velocities and motion velocities in three-dimensional space. Therefore, we set out to explore if the \texttt{H-star} and its host cluster have the same radial velocity. In the (a) panel of Fig.~\ref{Melotte25_Parameters_velocity}, it can be seen that the \texttt{H-star}'s radial velocity is beyond the range of the host's radial velocities, and even exceeds the nine times standard deviations of the host cluster's radial velocity. In addition, the two suspected intruders mentioned above also are beyond the range of the host's radial velocities. However, the radial velocity of a single observation for a star may be affected by the orbital rotation of a binary system or the pulsation of the variable source itself. To explore it, we obtained the time-series photometric data of the \texttt{H-star} and the suspected intruder marked by the black pentagram from the Transiting Exoplanet Survey Satellite (TESS) to analyze its light curve, with the \texttt{H-star} being named as TIC 411778231 and the suspected intruder marked by the black pentagram being called as TIC 47485851. Their light curves were displayed in Fig.~\ref{Lightcurve}. From this figure, we can see that TIC 47485851 has a significant photometric variation (black scatters), so it is a variable star, and its radial velocity is most likely not its radial velocity as a system. Therefore, this suspected intruder is most likely not an invasion star of Hyades. Moreover, for the suspected intruder star marked in purple, no photometric data were found for it from the TESS. However, from its position in the CMD, it is highly likely to be a binary star; thus its radial velocity is also most likely not its radial velocity as a system, and it may not be an intruder star of Hyades.

\begin{table*}[ht]
	\caption{The Parameters of the \texttt{H-star}}
	\centering
	
	\label{table:data}
	\begin{tabular}{c c c c}
		\hline\noalign{\smallskip}
		\hline\noalign{\smallskip}
		\hspace{0cm} Parameters & Description & Values & Units \\
		\hline\noalign{\smallskip}
		$\alpha$  & R.A. & 78.71~$\pm$~0.02 & degree  \\
		$\delta$  & decl. & 17.79~$\pm$~0.01 & degree  \\
		$\mu_{\alpha}$& Proper motion in R.A. & 54.28~$\pm$~0.02  & mas$\cdot$yr$^{-1}$\\
		$\mu_{\delta}cos\alpha$  & Proper motion in decl. & -27.43~$\pm$~0.01 & mas$\cdot$yr$^{-1}$ \\
		$\omega$  & Parallax & 17.63~$\pm$~0.02 & mas \\
		Rv  & The radial velocity & 13.62~$\pm$~1.36  & km$\cdot$s$^{-1}$\\
		V$_{X}$  & The space component velocity along x-axis  & -0.62~$\pm$~1.32  & km$\cdot$s$^{-1}$ \\
		V$_{Y}$  & The space component velocity along y-axis  & 229.95~$\pm$~0.13  & km$\cdot$s$^{-1}$\\
		V$_{Z}$ &  The space component velocity along z-axis &  12.82~$\pm$~0.28  & km$\cdot$s$^{-1}$ \\
		V &  The velocity with respect to the Sun & 21.29~$\pm$~0.87 & km$\cdot$s$^{-1}$ \\
		G  & G-band instrument's magnitude & 13.48~$\pm$~0.0028  & mag \\
		\hline\noalign{\smallskip}
	\end{tabular}
	\flushleft
	
\end{table*}

For the \texttt{H-star} (TIC 411778231), we found no significant photometric variations (red scatters) in Fig.~\ref{Lightcurve}. Meanwhile, Its radial velocity is 16.52 $\pm$ 2.97~km$\cdot$s$^{-1}$, 13.62 $\pm$ 1.36~km$\cdot$s$^{-1}$, and 16.62 $\pm$ 2.43~km$\cdot$s$^{-1}$ for Gaia, Gaia, and LAMOST observations, respectively. We note that the first two are published by Gaia~DR2 and Gaia~DR3, respectively, while the last one was determined by \citet{xian19} based on LAMOST~DR5. In a recent study of radial velocities for stars, \citet{tsan22} combined data from major surveys and determined the radial velocity of the \texttt{H-star} to be 15.84 $\pm$ 4.36~km$\cdot$s$^{-1}$. It is clear that its radial velocities measured by different telescopes at different times are roughly the same and stable. In short, there is no evidence that the \texttt{H-star} is a binary, either from its photometric analysis or from its radial velocity analysis.

Furthermore, we used the \texttt{Astropy SkyCoord} package to carry out all coordinate transformations and calculate their movement velocities in this work \citep{astr18}. For the package, we adopt the default values for the Galactocentric coordinate frame, namely: International Celestial Reference System coordinates (R.A., decl.) of the Galactic center = (266.4051$^{\circ}$, -28.936175$^{\circ}$); the Galactocentric Cartesian coordinates of the Sun: [$X_{\odot}$, $Y_{\odot}$, $Z_{\odot}$] = [8122, 0, 20.8]~pc, and the Galactocentric velocities of the Sun: [V$_{X_{\odot}}$, V$_{Y_{\odot}}$, V$_{Z_{\odot}}$] = [12.9, 245.6, 7.78]~km$\cdot$s$^{-1}$. Then, for all members with the radial velocities, their Galactocentric velocities [V$_{X}$, V$_{Y}$, V$_{Z}$] can be obtained through the \texttt{Astropy} package. Of course, their relative Galactocentric velocities also are available after removing the mean of their Galactocentric velocities, as shown in the (b) and (c) subfigures of Fig.~\ref{Melotte25_Parameters_velocity}. We noted that for the two suspected invading stars mentioned above, we did not calculate their relative Galactocentric velocities anymore, since their radial velocities may be false. We found that the relative Galactocentric velocities of the \texttt{H-star} are clearly different from that of its host cluster. Therefore, the \texttt{H-star} is most likely an intruder star in the Hyades cluster.

\subsection{Motion Orbit Analysis}

\begin{figure*}[ht]
	\centering
	\includegraphics[width=140mm]{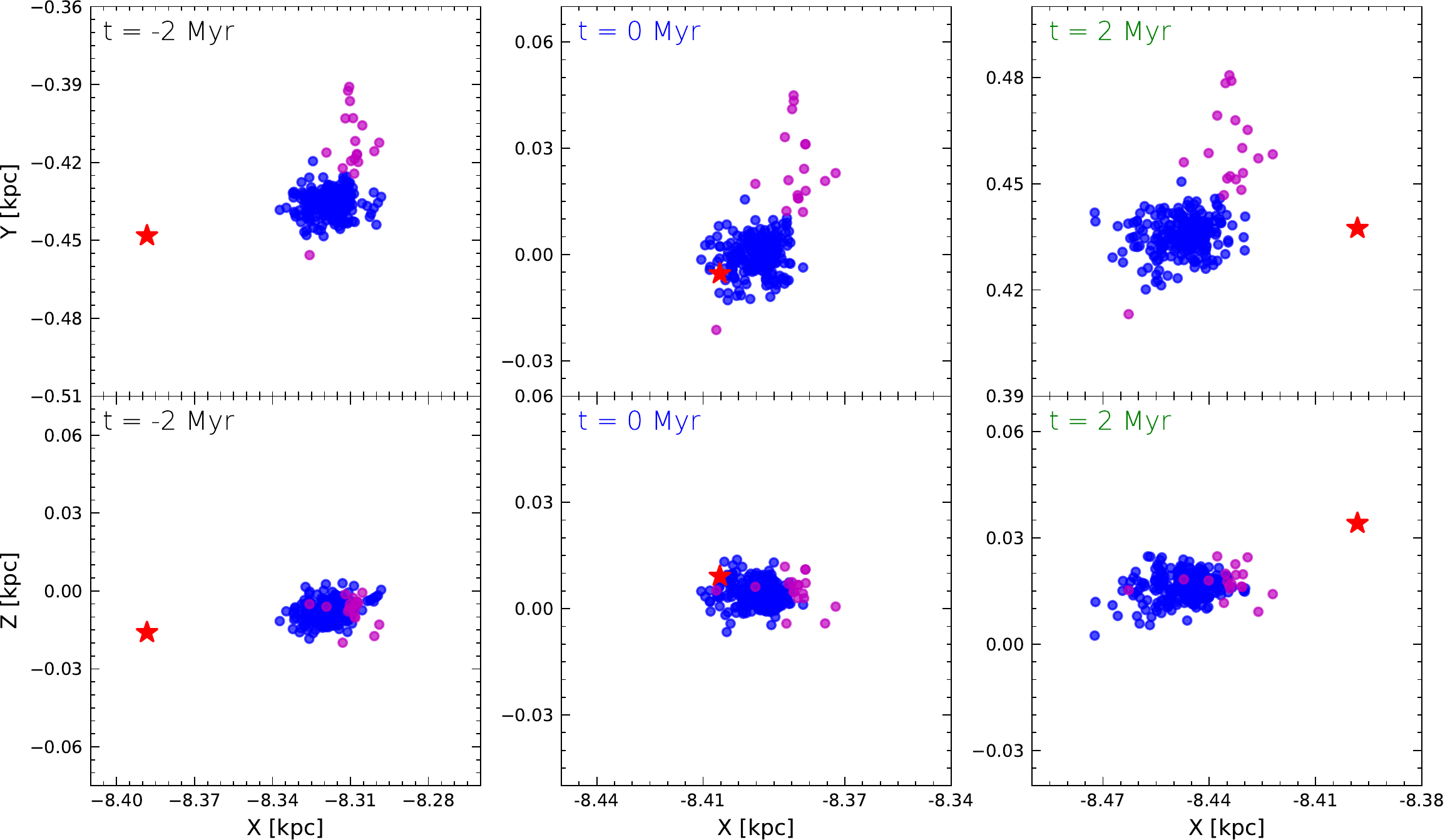}
	\caption{The projection distributions of the Hyades's member stars on different projection planes of the Galactic disk at some given moments (t = -2~Myr, 0~Myr, and 2~Myr). Colored dots are the same with Fig.~\ref{Melotte25_Parameters_position}.}
	\label{Melotte25_plane_XYXZ}
\end{figure*}

As mentioned above, the \texttt{H-star} may be an intruder member in the Hyades cluster. We in this subsection attempt to simulate its moving orbit within the \texttt{Galpy} framework. To explore it, we apply the \texttt{Python} module \texttt{galpy} \citep{bovy15} to integrate its movement track on the disk, based on the \texttt{galpy.potential} module, \texttt{MWPotential2014}. The axisymmetric potential module obtained by fitting a simple model to existing dynamical data of the Milky Way is composed of a bulge, a Miyamoto-Nagai disc \citep{miya75}, and a dark matter halo modeled with a Navarro-Frenk-White (NFW) potential \citep{nava97}. We note that only those members with the radial velocity in the Hyades can be used to simulate their movement orbits. Due to this reason, the adoptable number of cluster members is limited. Therefore, we need to seek the cluster members with the radial velocities as much as possible, while ensuring data homogeneity and reliability. Thus, we obtained such data (\textit{Sample c}), as described in Sect.~\ref{data}.

We analyzed the orbits of all these stars with the radial velocities within the \texttt{galpy} frame, shown in Fig.~\ref{Melotte25_Orbits}. In this figure, we showed their moving projection trajectories in the X-Y and X-Z projection planes, respectively. Their orbits on the Galactic disk are obtained by integrating the potential energy function forward and backward. To capture the details of the \texttt{H-star}'s intrusion into the Hyades and the \texttt{H-star}'s escape from this cluster, we arbitrarily integrate 60~Myr forward and 50~Myr backward, respectively. It can be seen that the orbit of the \texttt{H-star} is clearly different from that of the other members; see the dashed black and green lines in Fig.~\ref{Melotte25_Orbits}. We examine their projection distributions at some given moments (t = -2~Myr, 0~Myr, and 2~Myr) from the X-Y and X-Z planes, respectively, shown in Fig.~\ref{Melotte25_plane_XYXZ}. This suggests that if the current motion of the \texttt{H-star} does not change, it has probably entered the Hyades within the past 2~Myr and will escape from this cluster in the next 2~Myr.

\section{Discussion}

\subsection{Origin of the Intruder Star}

\begin{figure}[ht]
	\centering
	\includegraphics[width=50mm]{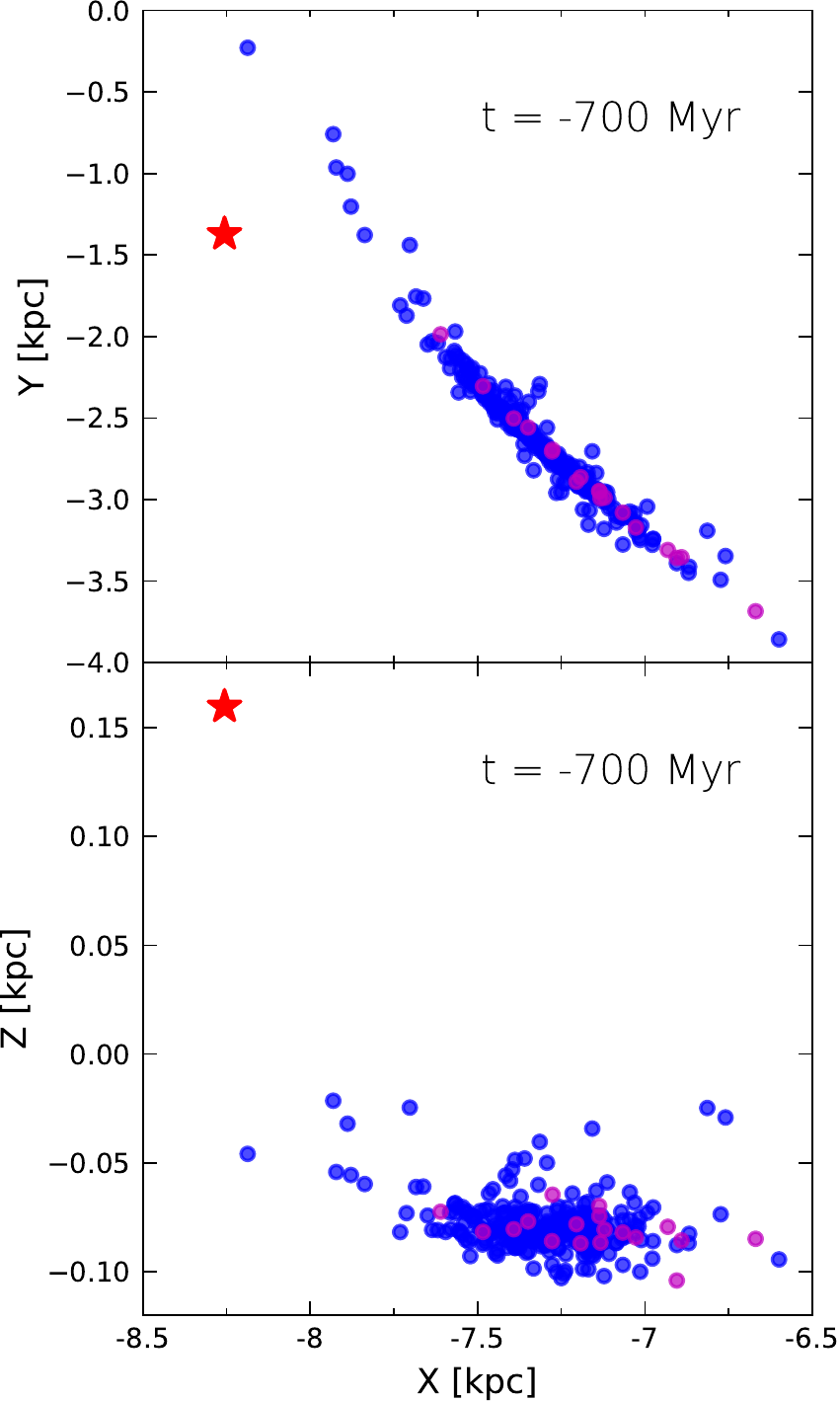}
	\caption{The projection distributions of the members of the Hyades on different projection planes of the Galactic disk at a given moment (t = -700~Myr). Colored dots are the same with Fig.~\ref{Melotte25_Parameters_position}.}
	\label{Melotte25_700Myr}
\end{figure}

We have already confirmed that the \texttt{H-star} is an intruder, both in terms of its kinematics and metal abundance. Furthermore, we would like to explore the possible origin of the \texttt{H-star}. Since its host has an intermediate age (about $>$~600~Myr), we ran the orbital model (\texttt{galpy}) up to 700 Myr ago to see if the intruder was mixed with members of Hyades at the point of its birth.
Fig.~\ref{Melotte25_700Myr} shows the distribution of the \texttt{H-star} and the Hyades' member stars on the X-Y and X-Z projection planes 700 years ago. We concluded from this figure that the \texttt{H-star} does not originate from the Hyades. So from where might it originate?
 
We can first draw an analogy between the \texttt{H-star} and runaways star \citep{bhat22}. There are two ways to produce a scenario similar to that of the runaways star. The first is a supernova explosion leading to the production of such a runaways star \citep[e.g.][]{hill83, burr95}, but this scenario usually occurs only in early-type stars. The \texttt{H-star} is the late M-type star. The second is a dynamical ejection scenario, i.e., the gravitational interaction of multiple stars leading to the production of a runaways star \citep[e.g.][]{hut83, hoff83}. And this scenario appears mostly in dense young clusters. The runaway stars formed in the dynamical ejection scenario must have been ejected very soon after their formation \citep{hoog01}. In both cases, the \texttt{H-star} more likely originated from other groups by dynamical ejection. Although its vertical velocity is actually more consistent with that of the thin disk group \citep{brat15}, it may be a member of the thick disk group \citep{mira16,zhuh21} due to its low metallicity. In summary, we speculated that it might originate from a disintegrated cluster or the Galactic field.

\subsection{Variable star type}

By the light curve of the star marked by the black pentagram in Fig.~\ref{Melotte25_Parameters_position}, we characterize it as a variable star. However, it is very important to determine what type of variable star it belongs to, which helps to explain why its radial velocity is not its systematic radial velocity. Therefore, we will simply explore its variable star type here. 

\begin{figure}[ht]
	\centering
	\includegraphics[width=76mm]{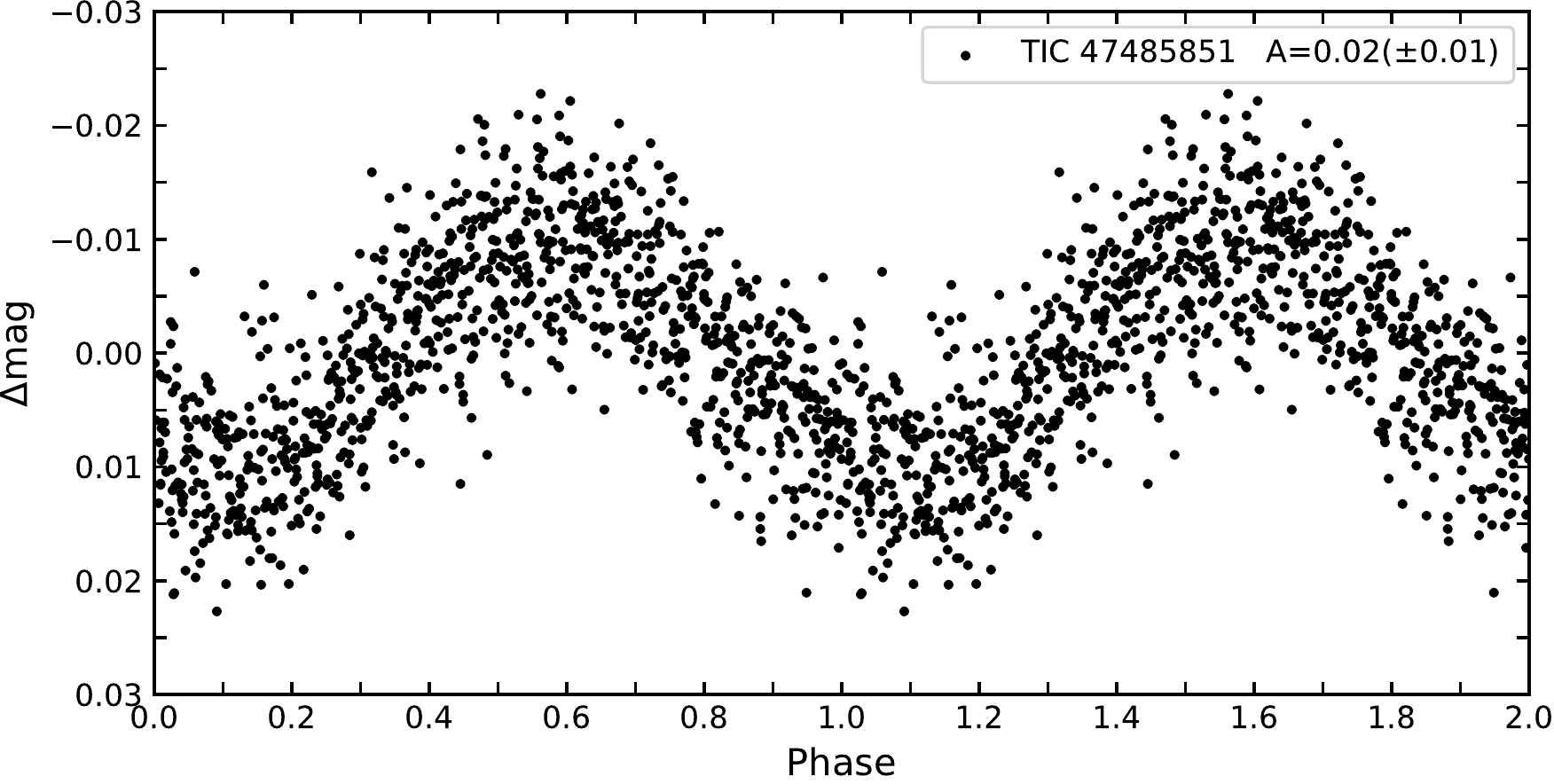}
	\caption{The phase diagram for the star (also TIC 47485851) marked by the black pentagram in Fig.~\ref{Melotte25_Parameters_position}.}
	\label{Phase}
\end{figure}

We use the amplitude, period, and effective temperature of the variable star to classify its variable star type. With the \texttt{python} package \texttt{LightKurve} \citep{ligh18}, we can obtain its frequency by taking the Fourier transform of its light curve. Its Fourier amplitude spectra show that it has only one significant frequency. Thus, we obtained its period by frequency inversion, a period of about 1.293 days. We then obtained its phase-folded diagram, folded by the period, shown in Fig.~\ref{Phase}. Due to the large dispersion of the phase-folded light curve, we applied the sliding average method to extract its light variation amplitude. The method is that we form a group of 60 data points and slide them backward in order to get a total of N-60 groups. Then, we can calculate the mean value of photometric magnitude in each group separately. Finally, we get its light variation amplitude according to the maximum and minimum average photometric magnitudes, the amplitude of about 0.02~($\pm$0.01)~mag. Its effective temperature (log(T$_{eff}$)) is mainly in the range of [3.466, 3.785], according to the parameters in the literature \citep{tonr18,muir18,Baiy19,ande19,stas19,cant20}.

According to the classification criteria \citet{Aert10} provided for variable stars, this star is likely to belong to the Cepheid variable star based on the parameters above. Then, it has likely a pulsation velocity up to 30 km$\cdot$s$^{-1}$. This means that its radial velocity (76.93~km$\cdot$s$^{-1}$~$\pm$~6.15~km$\cdot$s$^{-1}$ provided by Gaia DR3), as a member of Hyades, is not its systematic radial velocity, since this velocity has deviated from the cluster's radial velocity (about~40~km$\cdot$s$^{-1}$). Therefore, its radial velocity is most likely the sum of its systematic radial velocity and its pulsation velocity.

\section{Summary}

While studying the morphological evolution of open clusters, we accidentally detected an intruder star in the Hyades cluster. By comparing the kinematic features, metal abundances, and orbits of the \texttt{H-star} with those of its host, we confirmed that the \texttt{H-star} is a low-speed, late-type intruder for its host.

In detail, we analyzed the light curve of the \texttt{H-star} using the time-series photometric data from TESS, determining that it is not a variable star, and verifying the reliability of its radial velocity. We further investigated the significant inconsistency between the radial velocities of the \texttt{H-star} and its host cluster. Based on the literature data, we also found significant differences in the metal abundances between the \texttt{H-star} and its host. Therefore, it was found that the \texttt{H-star} is a low-velocity, low-metallicity star. Moreover, we discussed its possible origin in light of the orbital analysis of the \texttt{H-star} and its host in the \texttt{galpy} framework. By the distribution of the \texttt{H-star} and the host cluster on different projections of the Galactic disk at given moments, we found that the \texttt{H-star} may have invaded the cluster within the past 2 Myr. Meanwhile, if the current trajectory of the \texttt{H-star} remains unchanged, it will escape its host within the next 2 Myr.

In the analysis of the relationship between the \texttt{H-star} and the Hyades cluster, two suspicious intruders are found in the Hyades cluster. One of them is most likely a binary star and the other a variable star. In particular, for the former, we identified it as a binary based on its position in the CMD, since it has no available time-series photometric data. For the latter, we tentatively identified it as a Cepheid variable star according to its period, effective temperature, and amplitude.

For these two suspected intruders, we inferred that their current radial velocities may not be the radial velocities of their systems themselves in view of their variable star identity. In this case, we concluded that they are not intruders of Hyades.

\acknowledgments
We are grateful to an anonymous referee for valuable comments that have improved the paper significantly. This work was supported by the National Natural Science Foundation of China under grants (U2031209, U2031204, 12073070, and 11733008), the science research grants from China Manned Space Project with NO.CMS-CSST-2021-A08, and the National Key R\&D Program of China with No. 2021YFA1600403 and CAS `Light of West China' Program. We would also like to thank Ms. Chunli Feng for touching up the language of the article. This study has made use of the Gaia DR2 and Gaia DR3, operated by the European Space Agency (ESA) space mission (Gaia). The Gaia archive website is \url{https://archives.esac.esa.int/gaia/}. The metallicity parameters this work adopted are based on the data acquired through the Guoshoujing Telescope. Guoshoujing Telescope (the Large Sky Area Multi-Object Fiber Spectroscopic Telescope; LAMOST) is a National Major Scientific Project built by the Chinese Academy of Sciences. LAMOST is operated and managed by the National Astronomical Observatories, Chinese Academy of Sciences. The TESS data presented in this paper were obtained from the Mikulski Archive for Space Telescopes (MAST) at the Space Telescope Science Institute (STScI). STScI is operated by the Association of Universities for Research in Astronomy, Inc., under NASA contract NAS5-26555. Software: \texttt{Astropy} \citep{astr18}, \texttt{Galpy} \citep{bovy15}, \texttt{LightKurve} \citep{ligh18}, and \texttt{TOPCAT} \citep{tayl05}.



\end{document}